# Selectively addressing plasmonic modes and excitonic states in a nanocavity hosting a quantum emitter.

*Alberto Martín-Jiménez[1], Óscar Jover[1,2], Koen Lauwaet[1], Daniel Granados[1], Rodolfo Miranda[1,2], Roberto Otero[1,2]*

[1]IMDEA-Nanoscience Center, 28049 Madrid, Spain
[2]Dep. de Física de la Materia Condensada & IFIMAC, Universidad Autónoma de Madrid, 28049 Madrid, Spain

**Understanding and controlling the interaction between the excitonic states of a quantum emitter and the plasmonic modes of a nanocavity is one of the most relevant current scientific challenges, key for the development of many applications, from quantum information processing devices to polaritonic catalysts. In this paper we demonstrate that the tunnel electroluminescence of $C_{60}$ nanocrystals enclosed in the plasmonic nanocavity between a metallic surface and the tip of a Scanning Tunnelling Microscope, and isolated from the metal surface by a thin NaCl film, can be switched from a broad emission spectrum, revealing the plasmonic modes of the cavity, to a narrow band emission, displaying only the excitonic states of the $C_{60}$ molecules by changing the bias voltage applied to the junction. Plasmonic emission is found in the same voltage region in which the rate of inelastic tunnel transitions is large and, thus, vanishes for large voltages. Excitonic emission, on the other hand, dominates the spectra in the high-voltage region in which the inelastic rate is low, demonstrating that the excitons cannot be created by an inelastic tunnel process. These results point towards new possible mechanisms to explain the tunnel electroluminescence of quantum emitters and offer new avenues to develop electrically tuneable nanoscale light sources.**

## Introduction

The radiation from systems composed of a quantum emitter (QE) and a nanocavity depends on the strength of the light-matter coupling. In the weak coupling regime, the emission of QE is modified by the Purcell effect in the energy range covered by the plasmon resonances[1,2]. In the strong coupling regime, on the other hand, plasmonic and excitonic modes mix, leading to polaritonic excitations with half-light, half-matter character[1,2]. In order to control and understand the transition between both regimes, it would be very convenient to address the plasmonic and excitonic modes separately within the same system. Such a goal is, however, difficult to achieve with conventional optical spectroscopies, since the incoming radiation interacts with both, cavity modes and QE optical transitions due to its diffraction-limited spatial resolution. A related information can be retrieved by studying the cavity modes of the empty cavity and the excitonic states of the QE outside of the cavity[3], but this ignores the possible modification in QE and cavity properties when they are brought together.

Scanning Tunnelling Luminescence investigations can potentially open a new way to study QE-cavity interactions in systems consisting of organic molecules separated from metallic surfaces by an atomically thin insulating film[4–18]. The space between the metal surface and the metallic tip can be considered as a tuneable plasmonic nanocavity, whose broad optical modes can be studied by collecting tunnel electroluminescence spectra in molecule-free areas[9,10,13]. On the other hand, the electroluminescence spectra after injection of current directly on top of the molecules display narrow features that can be attributed to the recombination of molecular excitons (including side-bands related to Raman-active vibrational modes)[4–15]. The excitation of the plasmonic modes by STM has long been understood as the result of inelastic tunnelling processes[19,20], but the situation is not so clear for the excitonic case: while many previous studies also consider that such excitation is caused by inelastic events during the tunnel process (via an intermediate plasmon or not)[5–7,10,13], other reports explain the exciton formation as the result of two correlated elastic tunnelling processes, one from the tip to an unoccupied molecular orbital, and a second one from an occupied molecular orbital to the metal underneath the insulating film[4,14,15]. Notice that if the mechanisms for plasmon and exciton creation were different, one might conceive of ways to excite only the plasmonic or only the excitonic modes of the QE+cavity system, offering unique insights into the physics of light-matter interactions at the nanoscale, and providing us with a new tool to design colour-tuneable nanoscale light sources. One attempt in this direction was recently published for the case of multilayer $C_{60}$ on Ag(111), where the current-induced limitation of the exciton lifetime enabled a progressive transformation of excitonic spectra into plasmonic spectra with increasing tunnel current. The transformation, however, is very gradual and never complete[14].

In this paper, we demonstrate that the mechanisms for excitonic and plasmonic tunnel electroluminescence in $C_{60}$ nanoislands separated from a Ag(111) surface by a thin (2-3 ML) NaCl film are indeed different, enabling us to address only the plasmonic or only the excitonic modes of the system by a suitable change of the bias voltage of just about 0.5 V. Our study is based on the simultaneous recording of light intensities and the rate of inelastic excitations for different voltages and photon energies[21]. We show that, for the relatively narrow window of photon energies and bias voltages for which the rate of inelastic transitions is sizeable (between 2 and 2.8 V), electroluminescence spectra display a broad emission shape (FWHM ~ 200 meV), ascribed to the plasmonic resonances of the nanocavity, corresponding to the well-known fact that plasmonic emission is caused by inelastic events during tunnelling[19,20]. On the contrary, when the bias voltage is chosen to lay outside the range of high inelastic tunnelling rates for higher voltages, excitonic emission, equivalent to that found in single-crystal $C_{60}$[22], sets in, and dominates the spectra up to the maximum measured voltages (about 5.5 V). This fact implies that the energy transfer from the tunnelling electron to the exciton cannot occur during the tunnelling process, and thus, excludes inelastic excitation mechanisms. Moreover, the small separation from the LUMO level to the Femi level of Ag(111) also renders implausible the traditional two-tunnel events mechanism. In view of these results, we propose a new mechanism in which a hot electron in a $C_{60}$ molecule injected by an elastic tunnel process, relaxes though scattering processes that promote an electron from the HOMO to the LUMO, creating the exciton which subsequently decays by the emission of a photon. Our results, thus, shed new light on the mechanisms for excitonic light emission induced by tunnel currents, and open the fascinating possibility of choosing the excitation channel for light emission in systems composed of a QE and a nanocavity, which can be exploited for the fabrication of tuneable light sources in the nanoscale.

## Results and discussion

**Morphology and electronic structure of $C_{60}$ islands on NaCl/Ag(111).** Figure 1 shows the typical morphology of a $C_{60}$ nanocrystal supported on a thin NaCl film grown on Ag(111). The nanocrystals had characteristic truncated triangular shapes with an average lateral size of ~50 nm, with straight edges, and an apparent height of ~2 nm, that corresponds to a vertical stacking of two layers of $C_{60}$ fullerenes. Also, the corrugation of the NaCl template was visible through the top-most layer of molecules, as similarly reported for $C_{60}$ nanocrystals on NaCl-covered Au(111)[16,18,23]. Figure 1 b) shows a magnified view of the surface of the nanocrystal of a), displaying hexagonal self-assembly, with a nearest neighbour distance between molecules of 0.95 Å. Depending on the absorption configuration, the $C_{60}$ molecules displayed one-, two-, or three-lobed structures[23].

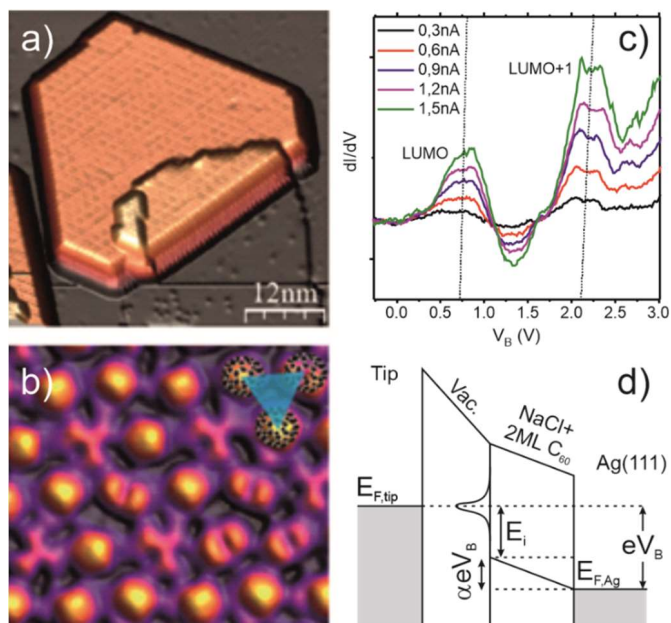

Figure 1. **Morphology and electronic structure of $C_{60}$ nanocrystallites.** a) STM topography image of a 2ML height C60 nanocrystal partially nucleated on top of 2 and 3ML of NaCl (2.5 V; 30 pA; 60 nm x 60 nm). b) Molecular self-assembly of the topmost layer of the nanocrystal shown in b. (1.5 V; 60 pA; 10 nm x 7 nm). Each molecular structure corresponds to a different adsorption geometry of the C60 molecules. c) dI/dV spectra recorded with different stabilization current setpoints. The LUMO and LUMO+1 orbitals are clearly visible in these spectra, but their voltage position depends on the tunneling conditions. d) Schematic diagram illustrating the relation between the applied bias voltage for the peak corresponding to a molecular orbital, the voltage drop between the surface of the crystallite and the NaCl interface, and the position of the molecular orbital with respect to its local electron chemical potential.

We have also characterized the electronic structure of the $C_{60}$ molecules by acquiring $dI/dV$ curves (Figure 1c), which show two significant features above the Fermi level at about 0.7 and 2.1 eV respectively, which we attribute to the LUMO and LUMO+1 orbitals of $C_{60}$. However, the exact position of these peaks depends on the tunnelling conditions, as should be the case due to the insulating nature of the NaCl spacer and the underlying layer of $C_{60}$ molecules[24]. The presence of this insulating film underneath the $C_{60}$ molecules on the surface implies that a fraction $\alpha$ of the applied voltage between the tip and the Ag(111) substrate falls between the crystallite surface and the NaCl/Ag interface, being $\alpha$ the ratio between the total capacitance of the junction and the capacitance of the insulating film (see Supplementary Information for more details). The Fermi level of the $C_{60}$ molecules at the surface of the nanocrystallites would thus be shifted by $\alpha eV_B$ with respect to the Fermi level of the Ag(111) substrate, and all the molecular levels will be shifted accordingly. Thus, the voltages at which the dI/dV peaks appear ($V_{B,i}$) will be related to the energies of the molecular orbitals at zero bias ($E_i$) by $eV_{B,i} = E_i + \alpha eV_{B,i}$ (see Figure 1d), or $eV_{B,i} = E_i/(1-\alpha)$. Changes in the tunnelling conditions, that is, tip-sample distances, lead to modifications in the total capacitance of the junction and $\alpha$ which, in turn, changes the voltages at which the peaks appear in the dI/dV curves, as observed in Figure 1c. Notice, however, that in the range of tunnelling parameters explored here, the shifts are small, implying that $\alpha$ is almost constant. We can obtain an estimation for $\alpha$ by comparing the energy separation between the LUMO and LUMO+1 orbitals in bulk $C_{60}$ obtained by inverse photoemission[25] and in our experiments, yielding a value of about $\alpha = 1 - \Delta E_{IPS}/\Delta V_{STM} = 0.23$.

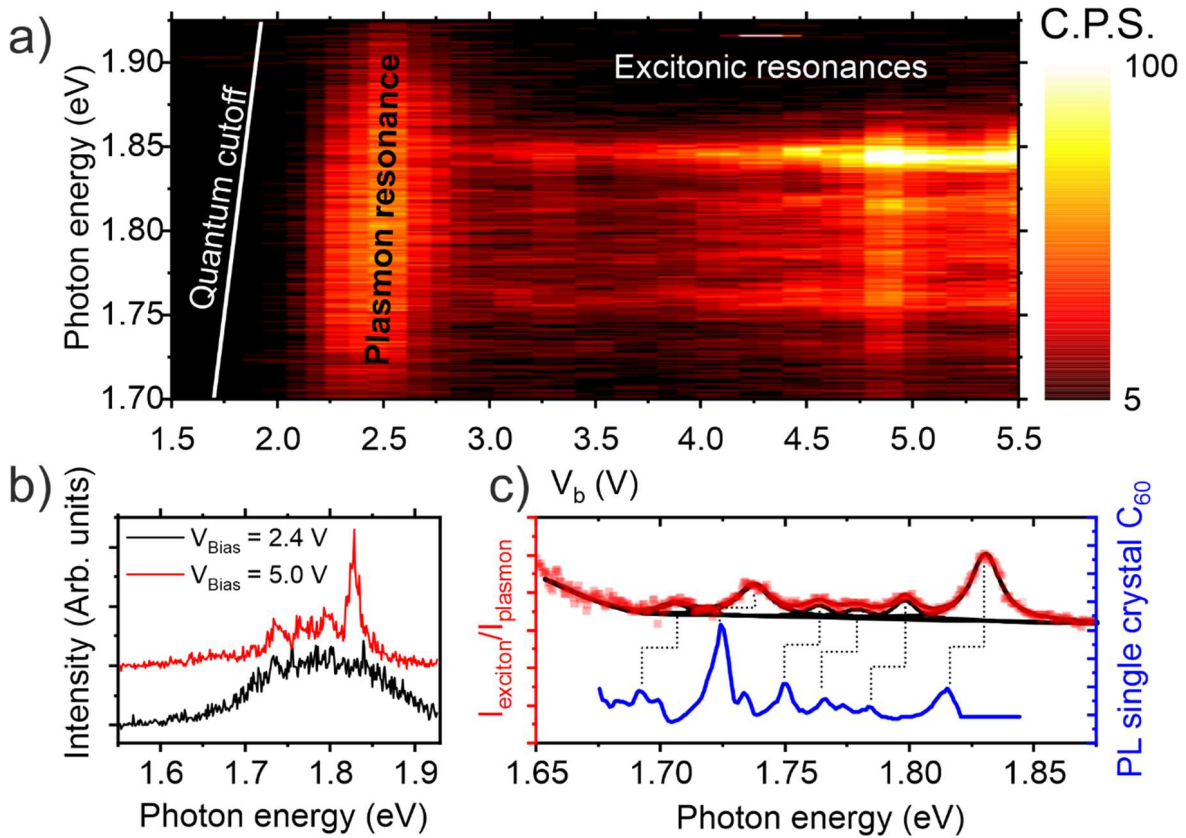

Figure 2. **Switching from pure plasmonic to pure excitonic emission**. a) Dependence of the electroluminescence spectra with the applied bias voltage for a tunnel current of 600 pA. The transition from a purely plasmonic to a purely excitonic spectrum is clearly visible between 2.8 and 3V. The expected quantum cutoff line is marked in white. b) Examples of individual spectra recorded at tunnelling conditions

where only the plasmonic or only the excitonic emission is observed. c) Comparison between the average electroluminescence spectrum at every voltage above 3 V (red dots) and the photoluminescence of defect-free bulk $C_{60}$ crystals[22] (blue line). Black lines are Lorentzian fits to the average, normalized spectrum.

**Plasmonic to excitonic transition in the electroluminescence spectra of $C_{60}$ islands on NaCl/Ag(111).** Tunnel electroluminescence spectra are shown in Figure 2a as a function of the bias voltage. The data reveal that in a relatively narrow window of bias voltage from 2 to 2.8 V, the spectra show a rather broad peak (FWHM~200 meV, see Figure 2b), characteristic of plasmonic emission. Contrary to simple metal systems, however, the voltage threshold for emission is shifted by about 0.3 V to higher voltages compared to the expected quantum cutoff condition ($\hbar\omega_{max} = eV_B$, white line in Figure 2a). The intensity of the plasmonic peak decreases very rapidly for voltages larger than 2.8 V, and basically disappears for 3 V. From 2.8 V on, however, a new set of luminescence peaks appear in the same photon energy region, consisting of an intense peak at about 1.83 eV, and several side bands at lower photon energies, which are very similar to those reported for $C_{60}$ nanocrystals on 3 ML NaCl/Au(111)[16,18]. The new peaks are much narrower than the plasmonic resonance (FWHM~15 meV, see Figure 2b), suggesting an excitonic origin. Indeed, by taking the average of all the spectra recorded for voltages between 3 and 5.5 eV and normalizing to the intensity of the plasmon (to take the Purcell enhancement into consideration), the resulting spectrum can be fitted by Lorentzian curves revealing all the features observed in photoluminescence of $C_{60}$ crystals[22] (see Figure 2c), with only a shift of about 25 meV to higher photon energies. This analysis demonstrates that the electroluminescence spectra in this range of bias voltages originates from the recombination of the same excitons observed in bulk $C_{60}$ crystals, including the side bands from the Raman-active vibrational levels.

We now wish to compare our results on the voltage dependence of the electroluminescence spectra with the rate of inelastic events, which can be retrieved from the $I(V)$ curves[21]. We have previously demonstrated that the consideration of such a rate allows for a proper understanding of the distance dependence of plasmonic resonances between the tip and a metallic sample[21], and of the functional form of the relation between overbias emission and electronic temperature[26]. To apply the same method to a system with an insulating film, however, requires some modification of the procedure. First, it should be noted that any inelastic event relevant for the processes considered here can only take place while the electron tunnels from the tip to the molecules at the surface of the crystallite (see Supporting Information for an extended discussion on this point). According to the scheme in Figure 1d and our previous discussion, the voltage drop in the vacuum tunnel process is just $(1 - \alpha)V_B$. The initial and final states of an inelastic tunnel process in which an excitation of energy $\hbar\omega$ is created must be separated by this energy, such as the one marked in a blue line in the scheme to Figure 3a. For a given applied voltage $V_B$, all the tunnel processes that can contribute to such excitation are marked in light blue, being limited by processes whose final state is the Fermi level at the surface of $C_{60}$ crystallite on the lower energy side, and by tunnel processes starting from the Fermi level of the tip and finishing at an energy $(1 - \alpha)eV_B - \hbar\omega$ with respect the neutrality point of the crystallite surface on the high energy side. The total rate at which such excitation of energy $\hbar\omega$ can be created by tunnelling electrons with an applied voltage $V_B$, thus, must be proportional to

$$R_{inel}(\hbar\omega, V_B) \propto \int_0^{(1-\alpha)eV_B-\hbar\omega} dE \rho_T(E + \hbar\omega - (1-\alpha)eV_B)\rho_S(E)T \quad (1)$$

In this expression $\rho_T$ and $\rho_S$ are the densities of electronic states of tip and sample respectively, $T$ is the transmission factor and we have assumed that the temperature is low enough to consider the Fermi-Dirac distributions to be step-functions. On the other hand, a similar argument can be used to obtain the elastic tunnel current that flows between the tip and the sample, yielding the same expression with $\hbar\omega = 0$.

$$I_t(V_B) \propto \int_0^{(1-\alpha)eV_B} dE \rho_T(E - (1-\alpha)eV_B)\rho_S(E)T \quad (2)$$

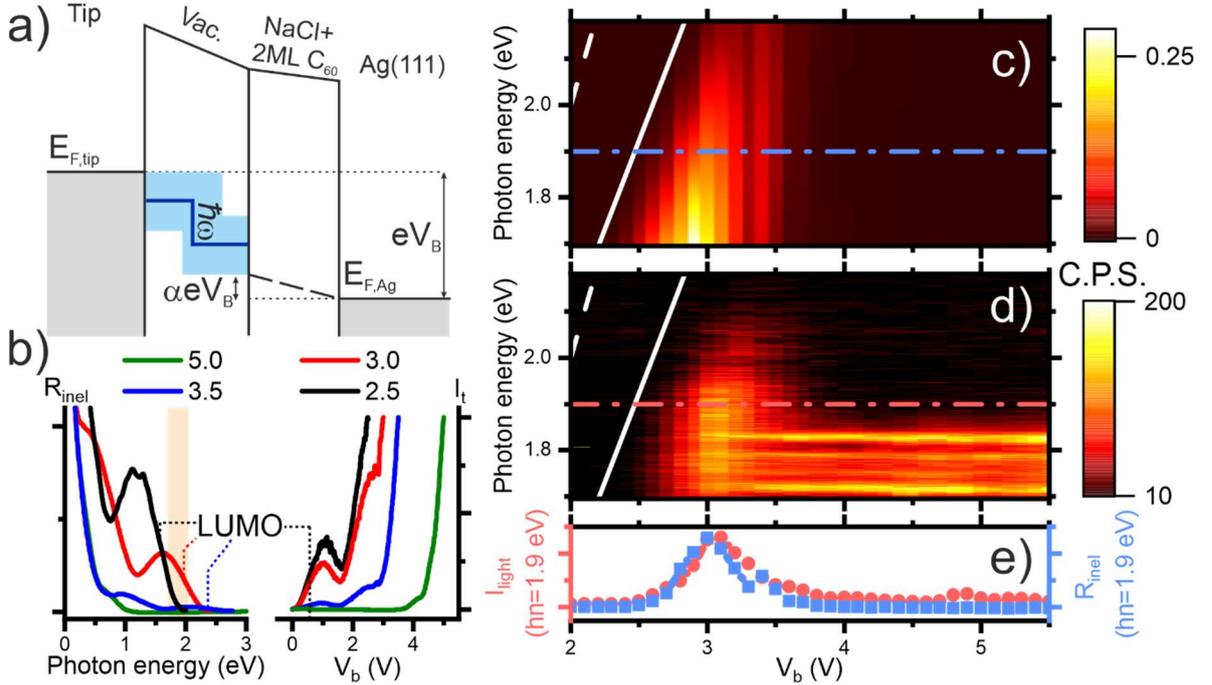

Figure 3. **The rate of inelastic transitions compared to electroluminescence spectra.** a) Schematic representation of the inelastic processes that can contribute to the excitation of a photon with energy $\hbar\omega$ in the presence of a dielectric spacer. b) Experimental rate of inelastic excitations (left panel) obtained from the I(V) curves (right panel) for different stabilization bias voltages. The orange area corresponds to the photon energy range studied in our experiments. c) Rate of inelastic excitation (linear color scale) for bias voltages between 2 and 5.5 V and photon energies 1.7 and 2.2 eV. The dashed line corresponds to the quantum cutoff condition without considering the dielectric layer, and the solid line corresponds to the correct quantum cutoff condition as determined by equation (3) by the condition $R_{inel}(\hbar\omega, V_B) = 0$. d) Electroluminescence signal (linear color scale) in the same photon energy and stabilization bias ranges than (c). Both quantum cutoffs are also shown with the same meaning than in (c). e) Comparison between the inelastic rate and electroluminescence profiles at $\hbar\omega = 1.9$ eV for different bias voltages. The profiles are also marked in (c) and (d) by the blue (rate) and pink (electroluminescence) horizontal dash-dot lines. In all the panels of this figure a tunnel current of 350 pA has been used.

The remarkable similarity between Equations (1) and (2) can be exploited to evaluate $R_{inel}(\hbar\omega, V_B)$ from the experimentally obtained $I(V)$ curves. Indeed, it is straightforward to realize that

$$R_{inel}(\hbar\omega, V_B) = I_t\left(V_B - \frac{\hbar\omega}{e(1-\alpha)}\right) \quad (3)$$

valid for $\hbar\omega < (1-\alpha)eV_B$. Equation (3) only differs from our previous analysis for purely metallic tunnel junctions[21] in the $(1-\alpha)^{-1}$ factor of the photon energy, which accounts for the fact that only a fraction $(1-\alpha)$ of the applied voltage between the Ag substrate and the tip can be invested in exciting photons. Thus, by recording the $I(V)$ curves at different stabilization voltages and applying Equation (3), we can obtain an experimental estimation of the rate at which inelastic transitions occur.

Figure 3(b) shows the application of this procedure for different stabilization voltages. The right-hand panel of Figure 3(b) displays the $I(V)$ curves. The onset of conductivity associated to the LUMO orbital at 0.7 eV can be clearly seen in all the curves (except for the one with a stabilization voltage of 5.0 V), followed by a region of negative differential conductance between 1 and 1.5 eV, and the onset of conductivity at 2.1 eV corresponding to the LUMO+1 orbital. With increasing stabilization voltages, the tip retracts to maintain the current setpoint constant. The current is then dominated by tunnelling between states with large vacuum decays, that is, with large energies. As a result, the tunnelling intensity becomes progressively dominated by voltages close to the stabilization voltage, and the relative contribution of lower-lying molecular orbitals is gradually reduced. In particular, the contribution of LUMO and LUMO+1 orbitals to the tunnelling current decreases with increasing stabilization voltage until it becomes negligible for stabilization voltages larger than about 3.5 V.

By applying Equation (3) we can obtain the rate of inelastic excitations as a function of the excitation energy as shown in the left panel of Figure 3(b). Since we do not observe a significant shift of the orbital positions with the stabilization voltage (Figure 1c), we consider $\alpha$ constant and equal to our previous estimation of 0.23. The general shape of the curves is now reverted in the energy axis, and the zero-rate value corresponds to the excitation energy $\hbar\omega = (1-\alpha)eV_B$, which becomes the new quantum cutoff condition. The window of photon energies that we explore by tunnel electroluminescence corresponds to the orange-marked area. For stabilization voltages below 2.5 V, the rates are low in this energy window, since the contribution of the LUMO orbitals covers a lower range of photon energies. Between 2.5 and 3.5 V, this contribution does overlap with the observed window of photon energies and, thus, we expect a relatively high contribution of inelastic events to the electroluminescence spectra, with electrons starting at tip states and finishing at the LUMO orbital of the $C_{60}$ molecules. For even higher voltages, however, the suppression of the contribution of LUMO and LUMO+1 orbitals to the total current leads to a rather strong decrease in the rate of inelastic excitations in the explored photon energy window. The vacuum decay of the LUMO and LUMO+1 orbitals is now negligible compared to the vacuum decay of higher-energy orbitals, and the inelastic tunnelling into these states is consequently suppressed.

Figure 3(c) collects all the rate curves measured for stabilization voltages between 2 and 5.5 V in the range of photon energies explored in the electroluminescence spectra. The (wrong) $\hbar\omega = eV_B$ cutoff condition is marked in a dashed line, while our new expected cutoff taking the dielectric film under consideration, $\hbar\omega = (1-\alpha)eV_B$, appears as a solid white line. A significant contribution of inelastic tunnel events is thus expected in the region bounded by the cut-off line and the high-bias suppression at about 3.5 V. We compare this expectation with simultaneously recorded electroluminescence spectra in Figure 3(d). Notice that, since we are using a different tip, the details of plasmonic resonance energies and widths might be somewhat different, but the general phenomenology agrees well with that reported in Figure 2(a): a broad plasmonic resonance can be observed for low voltages, and a set of narrow excitonic peaks with a maximum energy of 1.83 eV can be observed at higher voltages. Comparison with Figure 3(c), and between the spectral profiles for specific photon energies in Figure 3(e), reveals that the plasmonic resonance appears in the range of stabilization voltages and photon energies in which the rate of inelastic transitions is large, demonstrating that, as expected, the excitation of the plasmonic luminescence is induced by inelastic tunnelling. On the contrary, excitonic peaks appear in the region in which the rate of inelastic transitions is negligible, because of the weak overlap between tip electronic states and $C_{60}$ LUMO and LUMO+1 molecular orbitals. The creation of excitons, thus, is not related to inelastic tunnelling and must be originated by another process that takes place after injection of the hot electron in the $C_{60}$ molecule. We conclude that, in our QE+cavity system (QE=$C_{60}$ molecules, cavity=tunnel gap between the metallic tip and the Ag substrate), we can address independently the plasmonic or the excitonic modes simply by changing the bias voltage since the excitation mechanisms to plasmonic and excitonic luminescence are of a different nature.

**Implications for the excitation mechanism of excitonic luminescence from $C_{60}$ islands on NaCl/Ag(111).** The excitation of molecular luminescence deserves further discussion. In general, the mechanism for excitonic creation induced by a tunnel current is still under debate in the literature. Some authors have attributed the promotion of the molecules from the electronic ground state to the excited state to inelastic tunnelling events, either mediated by a plasmon resonance, or a in a more direct fashion[5–7,10,13]. Our results demonstrate that, at least for our system, this mechanism is not the one responsible for molecular luminescence. An alternative explanation is that the shift in the molecular orbitals due to the dielectric layer that we describe here might push the HOMO level above the Fermi level of the metal substrate, thereby enabling the tunnel of an electron away from the HOMO. Elastic injection of electrons into the LUMO orbital would lead thus to a situation with an electron and a hole in the same molecule which, upon electronic relaxation, can become the excited state of the molecule containing one exciton[4,14,15]. However, such mechanism does not match our experimental observations either. In particular, the LUMO level is only 0.6-0.7 eV above the Fermi level and, while it does shift with tunnelling conditions, the shifts are very small. At 3.5 V, where excitonic emission is clearly discernible, the LUMO level is only about 0.7 eV above the Fermi level. Even if the shift of the molecular orbitals with the bias voltage was not rigid (as we are assuming here) and the HOMO level would rise above the Fermi level of the sample in the voltage range explored in our investigations, its separation with the LUMO level would be at most 0.7 eV, too small compared with the exciton energy of 1.83 eV that we observe from the $C_{60}$ (and consideration of exciton binding energies can only make the emission energy smaller). Other processes involving the injection of electrons into the LUMO+1 or higher molecular orbitals followed by a recombination with holes in the LUMO, a hypothetical emptied HOMO level or other final state alternatives should lead to very different luminescence spectra than that recorded on bulk $C_{60}$ crystals, where the transition was explained as a HOMO-LUMO transition facilitated by the Herzberg-Teller effect. However, the similarity between the PL signal in bulk $C_{60}$ crystals and our data shown in Figure 2(c) renders this scenario involving different excitonic states unlikely.

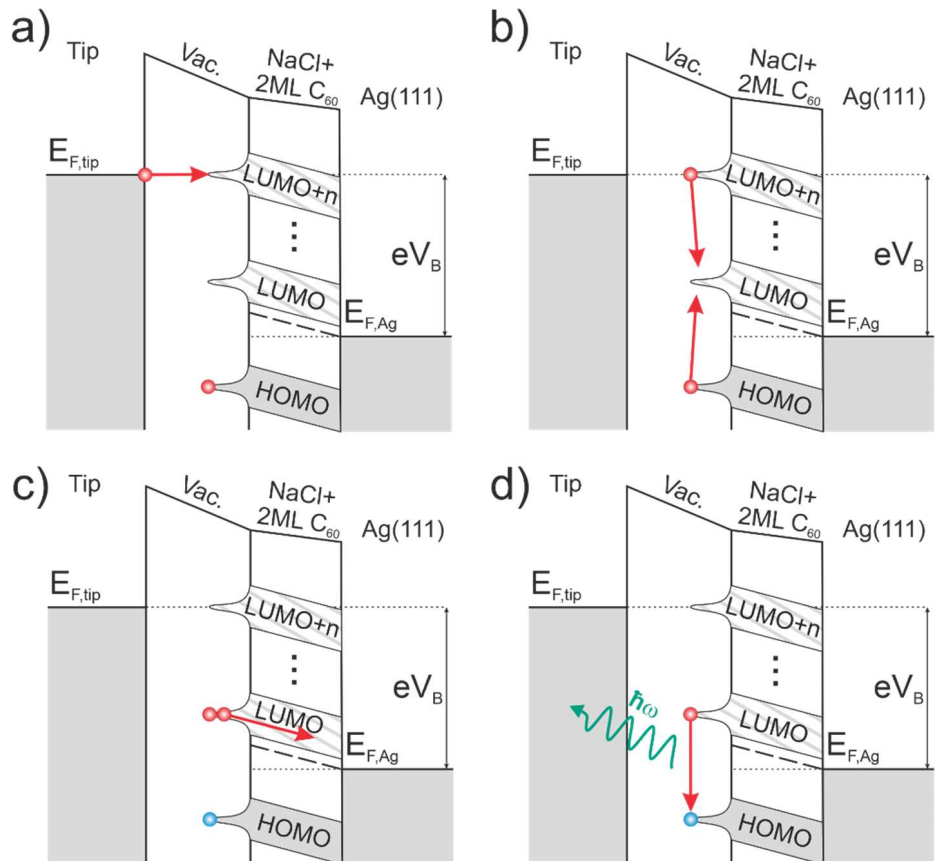

**Figure 4. Proposed hot electron mechanism for excitonic tunnel electroluminescence.** a) Initially an elastic tunnel event creates a hot electron in a high energy state, while the electron population of the remaining levels does not change. b) This hot electron will relax by scattering with other electrons in the occupied bands of the $C_{60}$ nanocrystal. If the energy of the hot electron is at least the energy of the LUMO level plus the exciton energy, then a relaxation in which the hot electron relaxes to the LUMO and an exciton is created becomes possible. c) The electron diffuses away from the excited $C_{60}$ molecules, leaving a neutral exciton indistinguishable from those created by optical excitations. d) The exciton decays radiatively through the emission of a photon.

Based on our data and the preceding discussion, we propose the following mechanism to explain our observation of excitonic emission induced by tunnel currents. A hot electron produced after elastic tunnel injection (Figure 4(a)) in a narrow unoccupied band higher than the LUMO can relax to a lower-lying band by scattering processes with electrons in the (also narrow) occupied bands originated from the HOMO and lower molecular orbitals. For these processes to occur, however, the electrons in the occupied bands must be able to absorb the excess energy of the hot electron, and thus, must be promoted to empty bands by the scattering process. The minimum transition energy corresponds to the difference between the LUMO and the HOMO energies, minus the electron-hole binding energy; that is, by definition, the exciton energy. Thus, provided that the hot electron has an energy of at least the exciton energy above the LUMO level, it can relax to the LUMO via creation of an exciton (see Figure 4(b)). If the exciton lifetime is larger than the inverse of the hopping rate of the hot electron, it will diffuse away before exciton recombination (Figure 4(c)), which, when it finally takes place, will only reflect the character of the neutral exciton (Figure 4(d)), explaining the similarity between our data and that of photoluminescence in bulk $C_{60}$ crystals. These considerations also allow for an estimation of the bias threshold to excitonic emission, which should satisfy $eV_{B,th} = \alpha eV_{B,th} + E_{LUMO} + E_{exciton}$, or

$$eV_{B,th} = \frac{E_{LUMO} + E_{exciton}}{1 - \alpha} \approx 3 \text{ eV}$$

This estimation is in good qualitative agreement with the data in Figure 2(a) and 3(d), although some excitonic emission can also be observed at some 0.2 V lower bias voltages, an effect that we attribute to the broadening of the LUMO orbital (the onset of which can be estimated at around 0.3 eV, as can be seen in Figure 1(c)).

To conclude, our systematic study on the dependence of electroluminescence and inelastic rates with the applied bias voltage of $C_{60}$ nanocrystallites on 2ML NaCl/Ag(111) has demonstrated that the mechanisms for plasmonic and excitonic radiation are radically different: whereas inelastic tunnel events induce the former process, the later originates from excitons which are created by relaxation of hot electrons injected at high energies. The difference in the excitation mechanisms enables us to choose our tunnelling parameters to promote one of either at will, thereby interrogating only the plasmonic or only the excitonic modes of a QE+cavity system. These results thus open new scenarios to understand molecular electroluminescence processes and can be exploited to design new colour-tuneable, nanoscale light sources.

## Methods

**Sample and tip preparation.** All the experiments were performed with an Omicron Low-Temperature Scanning Tunnelling Microscope (LT-STM), operated at 4.5 K, in Ultra-High-Vacuum (UHV) conditions (P ~ $10^{-11}$ mbar), and equipped with a custom-built light detection setup[21]. Ag(111) single crystals (MaTecK) were cleaned by repeated cycles of sputtering with 1.5 keV $Ar^+$ ions followed by thermal annealing at 500 K. The nanocrystallites were grown by deposition of $C_{60}$ fullerenes (Sigma Aldrich) on top of a NaCl covered Ag(111) sample. The NaCl template (2-3 mono-layers) was grown by sublimation of NaCl (Sigma Aldrich) onto the clean Ag(111) surface held at room temperature. Au tips were used in all the experiments. After electrochemical etching in a solution of HCl (37%) in ethanol, the tips were cleaned in UHV by sputtering with 1.5 keV $Ar^+$ ions.
**Data acquisition.** All the topography images and luminescence spectra were acquired in the constant current mode of the STM. The topography images were processed with the WSxM software. The luminescence spectra were recorded with an Andor Shamrock 500 spectrometer equipped with a Peltier cooled Charge Coupled Device (CCD). The CCD was operated at -85 C. The acquisition time of each spectrum in figure 2a and 3d was 100 seconds, and the spectra were recorded in the voltage range from 1.5 to 5.5 V in steps of 0.1 V.

## Acknowledgments

R.M. and R.O. acknowledge financial support from the Spanish Ministry for Science and Innovation (Grants PGC2018-098613-B-C21, PGC2018-096047-B-I00), the regional government of Comunidad de Madrid (Grant S2018/NMT-4321), Universidad Autónoma de Madrid (UAM/48) and IMDEA Nanoscience. Both IMDEA Nanoscience and IFIMAC acknowledge support from the Severo Ochoa and Maria de Maeztu Programmes for Centres and Units of Excellence in R&D (MINECO, Grants CEX2020-001039-S and CEX2018-000805-M). R.O. acknowledges support from the excellence programme for University Professors, funded by the regional government of Madrid (V PRICIT).

# Supplementary Information

# Selectively addressing plasmonic and excitonic modes in a nanocavity hosting a quantum emitter.

*Alberto Martín-Jiménez[1], Óscar Jover[1,2], Koen Lauwaet[1], Daniel Granados[1], Rodolfo Miranda[1,2], Roberto Otero[1,2]*

[1]*IMDEA-Nanoscience Center, 28049 Madrid, Spain*
[2]*Dep. de Física de la Materia Condensada & IFIMAC, Universidad Autónoma de Madrid, 28049 Madrid, Spain*

1- Effect of the dielectric layer on the tunnelling characteristic curves.

The existence of a dielectric layer separating the molecules at the surface of the crystallites from the metal substrate implies that only a fraction of the applied voltage drops between the STM tip and the C60 molecules, while the other part of the voltage shifts the charge neutrality level of the crystallite surface with respect to the Fermi level of the Ag(111) substrate. In order to estimate this fraction, we model the junction as a capacitor filled in part by vacuum and in part by an effective dielectric that includes the underlying C60 molecules and the NaCl film. The capacitance of the junction can thus be considered to result from the association in series of two capacitors, one filled with vacuum and the other one filled with the dielectric. We can thus state that

$$C_T V_B = C_{ins} V_{ins}$$

(S1)

where $V_B$ is the total bias voltage applied between the Ag(111) surface and the Au tip, $C_T$ is the total capacitance of the junction, $V_{ins}$ is the voltage that drops between the surface of the crystallite and the Ag(111) interface and $C_{ins}$ is the capacitance associated to the effective dielectric composed of the C60 layers behind the surface and the NaCl film. Thus, the voltage drop in the dielectric layer is simply proportional to the applied total voltage with a proportionality constant given by $\alpha = C_T/C_{ins} \leq 1$. $\alpha$ depends on the tip-surface separation, and thus, comparing curves recorded with different stabilization bias would in principle require to calculate $\alpha$ for both conditions. Within a dI/dV or I(V) curve, however, since the feedback loop is open, the distance between tip and surface remains constant and, thus, one can assume a single value of $\alpha$ for all the curve.

2- Vacuum versus dielectric tunnelling as possible luminescence sources.

Direct tunnelling from the tip to the Ag surface underneath must be a rare event, since the distance between the tip and the Ag surface is about 2.5 nm (the thickness of 2ML C60 plus 2ML NaCl) larger than between the tip and the surface of the molecule. On the other hand, a hypothetical second tunneling event from the surface of the crystallite to the Ag interface would occur at an effective bias voltage $\alpha V_B$. With our previous estimation of $\alpha$ and for the maximum voltages used in this experiment, the separation between the Fermi levels of crystallite surface and Ag interface can be of just 1.26 V, not enough to excite plasmons of excitons with energies of about 1.8 eV.